\documentclass[aps,prl,groupedaddress,showpacs]{revtex4}
\usepackage{graphicx}

\def\be{\begin{equation}}
\def\ee{\end{equation}}
\def\ba{\begin{eqnarray}}
\def\ea{\end{eqnarray}}
\def\bc{\begin{center}}
\def\ec{\end{center}}

\begin{document}

\title{Non-linear effects in the cyclotron resonance of a massless quasi-particle in graphene}

\author{S. A. Mikhailov}
\email[Electronic mail: ]{sergey.mikhailov@physik.uni-augsburg.de}

\affiliation{Institute of Physics, University of Augsburg, D-86135 Augsburg, Germany}

\date{\today}

\begin{abstract}
We consider the classical motion of a massless quasi-particle in a magnetic field and under a weak electromagnetic radiation with the frequency $\omega$. Due to the non-parabolic, linear energy dispersion, the particle responds not only at the frequency $\omega$ but generates a broad frequency spectrum around it. The linewidth of the cyclotron resonance turns out to be very broad even in a perfectly pure material which allows one to explain recent experimental data in graphene. It is concluded that the linear response theory does not work in graphene in finite magnetic fields.
\end{abstract}


\maketitle

Graphene is a recently discovered two-dimensional (2D) material \cite{Novoselov05,Zhang05} which has attracted much attention due to its non-trivial and very interesting physical properties, see reviews  \cite{Katsnelson07,Geim07a}. This is a monolayer of carbon atoms packed in a dense honeycomb lattice. The band structure of electrons in graphene consists of two bands (electron and hole) which touch each other at six corners of the hexagon-shaped Brillouin zone \cite{Wallace47}. If graphene is undoped and the temperature is zero, the lower (hole) band is fully occupied while the upper (electron) band is empty. The Fermi level goes through these six, so called Dirac points. Near these points the states of electron and holes are described by the effective Dirac equation with the vanishing effective mass of the quasiparticles, and their energy spectrum is linear, 
\be
{\cal E}_\pm({\bf p})=\pm Vp=\pm V|{\bf p}|.
\label{disp}
\ee
Here ${\bf p}=(p_x,p_y)$ is the electronic momentum, counted from the corresponding Dirac points and $V\approx 10^8$ cm/s is the material parameter. It is the massless energy dispersion (\ref{disp}) and the Dirac nature of graphene quasiparticles that result in its amazing physical properties which attracted so much attention in the past years.

If a charged massive particle (e.g. electron) is placed in a magnetic field ${\bf B}$, it begins to rotate in the plane perpendicular to the field ${\bf B}$ with the frequency $\omega_c=eB/mc$, determined by its mass $m$ and charge $e$. If such a rotating particle is irradiated by an external electromagnetic wave with the frequency $\omega$, it absorbs the radiation energy if the frequency of the wave is close to the cyclotron frequency, $\omega\simeq\omega_c$. The phenomenon is called the cyclotron resonance (CR) and is widely used in solid state physics for  characterization of material properties such as the charge carrier effective mass and the Fermi surface cross-section in metals and semiconductors \cite{LAK}. The linewidth of the CR absorption line $\delta\omega$ is determined by the scattering rate of 2D electrons and by the radiative decay rate, i.e. by losses of energy of the rotating charged particle due to the re-radiation of electromagnetic waves,  see e.g. \cite{Mikhailov04a}. In typical semiconductor 2D electron systems, e.g. in GaAs quantum well structures, the linewidth of the CR is small as compared to the CR frequency already in magnetic fields of order of 0.1 Tesla (see e.g. \cite{Kukushkin06}).

The magneto-optical conductivity of graphene and its cyclotron response have been studied within the linear response theory in a number of publications \cite{Peres06,Gusynin06a,Gusynin07b,Barlas08,Morimoto08}. Intensive experimental studies of the CR in graphene (as well as other electrodynamic phenomena) have been hampered until recently by the absence of graphene flakes of sufficiently large area. With the progress of technology, however, graphene samples of sufficiently large dimensions are becoming available now \cite{Nair08}, so the growth of experimental activity on the graphene electrodynamics is to be expected. First experimental results on the CR in single-layered graphene have been already published in Refs. \cite{Jiang07,Deacon07} (for the CR data in bilayer graphene and in thin graphite layers see Refs. \cite{Henriksen08,Sadowski06}). In Ref. \cite{Jiang07} the infrared spectroscopy has been used and in Ref. \cite{Deacon07} the photoconductive response of graphene has been measured. The  characteristic feature of both CR experiments \cite{Jiang07,Deacon07} is that the CR line was found to be extremely broad. Although the experiments have been done in very strong magnetic fields (up to 18 T), the quality factor of the CR line $B_c/\delta B_c\simeq\omega/\delta\omega$  only slightly exceeded unity. Following the traditional interpretation of the CR linewidth one had to conclude that in Refs. \cite{Jiang07,Deacon07} one dealt with extremely disordered graphene samples. As will be shown below, however, the cyclotron resonance of massless quasi-particles (\ref{disp}) has very unusual physical properties, and {\em the CR-line can be very broad even in perfectly pure graphene}. 

The most straightforward way to demonstrate the main idea of this work is to consider the classical motion of a quasi-particle with the spectrum (\ref{disp}) in a uniform magnetic field ${\bf B}=(0,0,B)$ and in the presence of an external electromagnetic wave. The equation of motion reads
\be
\frac{d{\bf p}}{dt}=-\frac {e}c{\bf v}\times {\bf B}-e{\bf E}(t), \ \ {\bf v}=V\frac{\bf p}{p}.
\label{Newton}
\ee
If the external electric field ${\bf E}(t)$ is zero, the particle rotates in the magnetic field with the energy-dependent cyclotron frequency $\omega_c=eBV/pc=eBV^2/c{\cal E}$, where the energy ${\cal E}=Vp$ is the integral of motion (e.g. \cite{Landau2}, \S 21). If the external electric field is switched on, one usually assumes that the particle absorbs the electromagnetic wave energy at the cyclotron frequency $\omega_c$. This is perfectly true for electrons with the parabolic spectrum, but for the massless graphene quasi-particles (\ref{disp}) the situation turns out to be much more complicated and more interesting. 

\begin{figure}
\includegraphics[width=8.5cm]{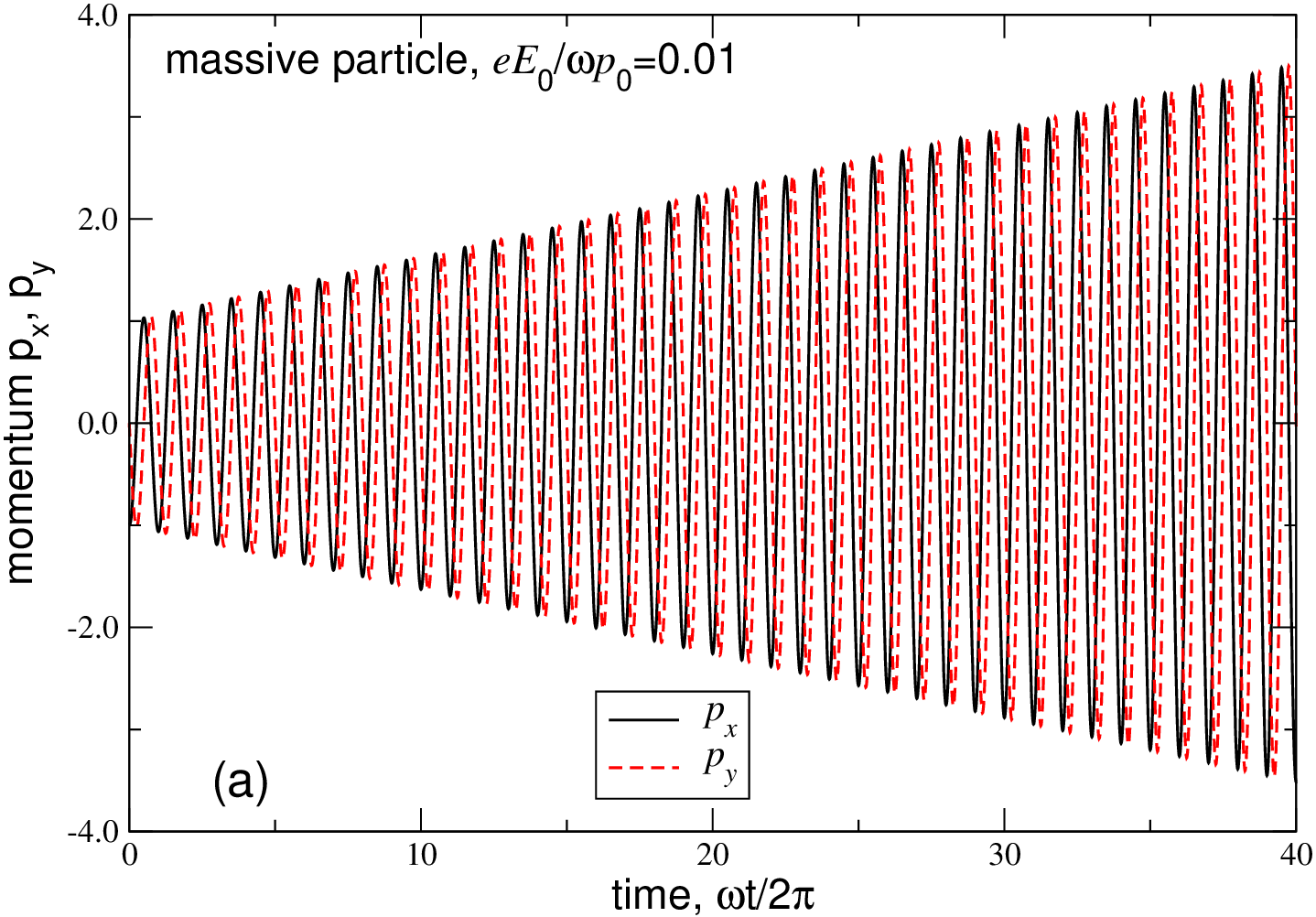}\hfill 
\includegraphics[width=8.5cm]{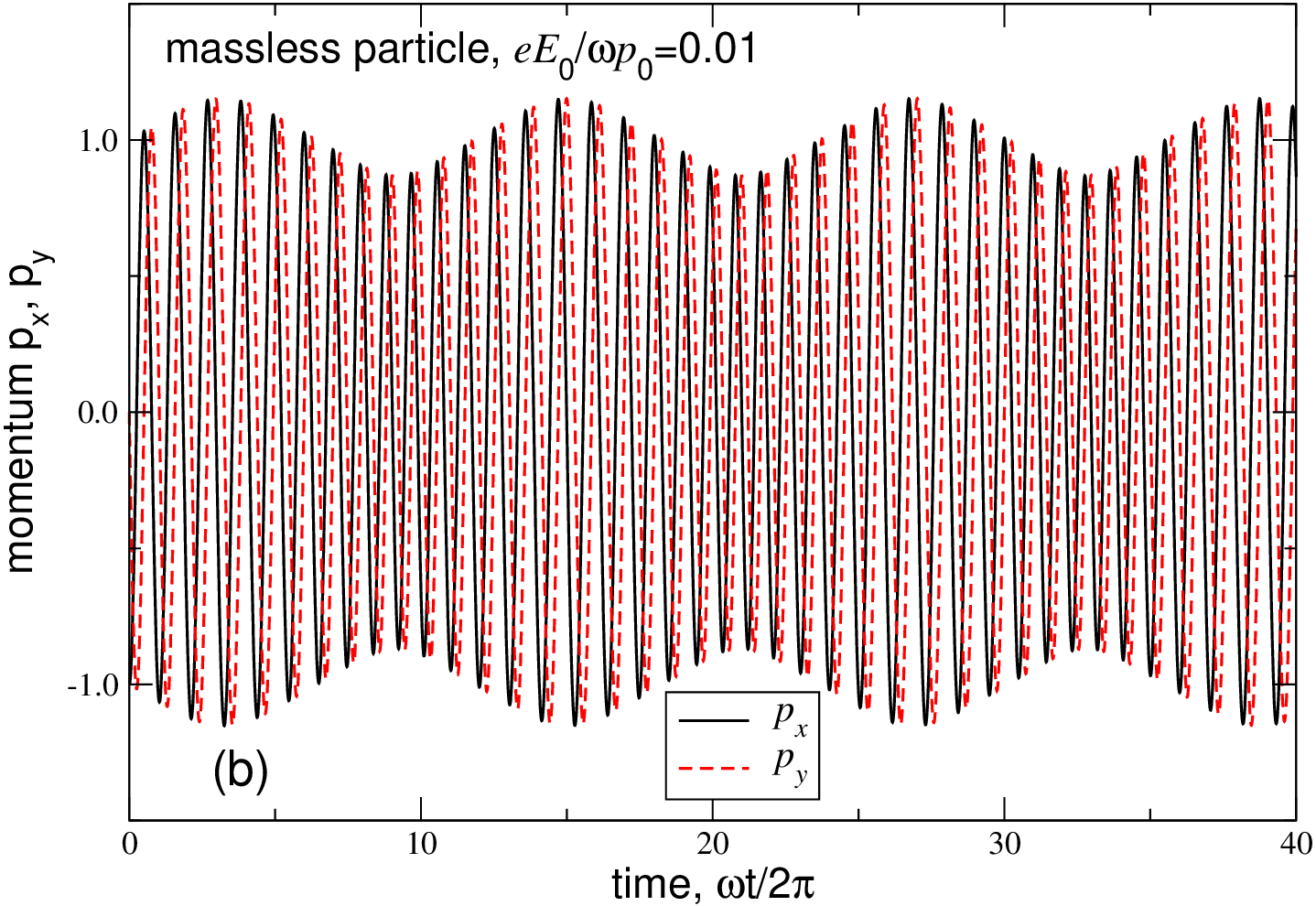}
\caption{\label{traj} (Color online) Time dependencies of the momentum of the massive {\bf (a)} and the massless {\bf (b)} quasi-particles under the CR conditions. 
}
\end{figure}

In Figure \ref{traj} we show results of the numerical solution of Eq.  (\ref{Newton}) with the initial conditions ${\bf p}|_{t=0}={\bf p}_0$. We assume that the radiation frequency $\omega$ coincides with the cyclotron frequency $\omega_{c}$ and that the external electric field ${\bf E}_0(t)=E_0(t)(\cos\omega t,\sin\omega t)$ is circularly polarized in the direction coinciding with the sense of the cyclotron rotation of the particle. The spectrum of the incident radiation $S_{inc}(\Omega)$ thus consists of only one spectral line, $S_{inc}(\Omega)\propto\delta(\Omega-\omega)$, and the particle is under the perfect CR conditions. For the sake of clarity we further neglect all dissipative processes such as the scattering of electrons by impurities and phonons, as well as the radiative decay (\cite{Landau2}, \S 67). We also assume that the external electric field $E_0$ is weak, ${\cal F}\equiv eE_0/\omega p_0\ll 1$,  meaning that the energy absorbed by the particle during one oscillation period is small as compared to its average energy. Usually this corresponds to the linear response regime. 

Figure \ref{traj}a shows the known results for a standard (massive) particle (in this case ${\bf v}={\bf p}/m$ in (\ref{Newton}) and $\omega_{c}=eB/mc$). Under the chosen conditions, the massive particle is {\em always} in resonance with the external field, it continuously absorbs the radiation energy, and the absolute value of its momentum $p$ linearly grows with time. Rotating with the frequency $\omega$ the massive particle re-emits (scatters) the radiation of the incident wave at the same frequency $\omega$, and the spectrum of the scattered waves $S_{scat}(\Omega)$ is also proportional to $\delta(\Omega-\omega)$.

The motion of the massless particle is essentially different. Its cyclotron frequency $\omega_{c}=eBV/pc=eBV^2/{\cal E}c$ depends on its energy. Initially, when the external field is switched on, the particle is also in resonance with the external radiation, $\omega=\omega_{c}=eBV/p_0c$, and starts to get energy from the  wave. But, as soon as its energy increases, $p>p_0$, it gets out of the resonance and ceases to absorb the radiation energy. Its energy ${\cal E}=Vp$ and the absolute value of the momentum $p$ decreases and the particle again turns out to be in the resonance with the wave. Then the process repeats itself, Fig. \ref{traj}b, and the particle energy oscillates in time with the period depending on the external field amplitude $E_0$. The time dependence of the momentum of the massless particle is {\em not a harmonic function} with only one Fourier component $\Omega=\omega$. It also contains other (higher and lower) harmonics.

\begin{figure}
\includegraphics[width=8.5cm]{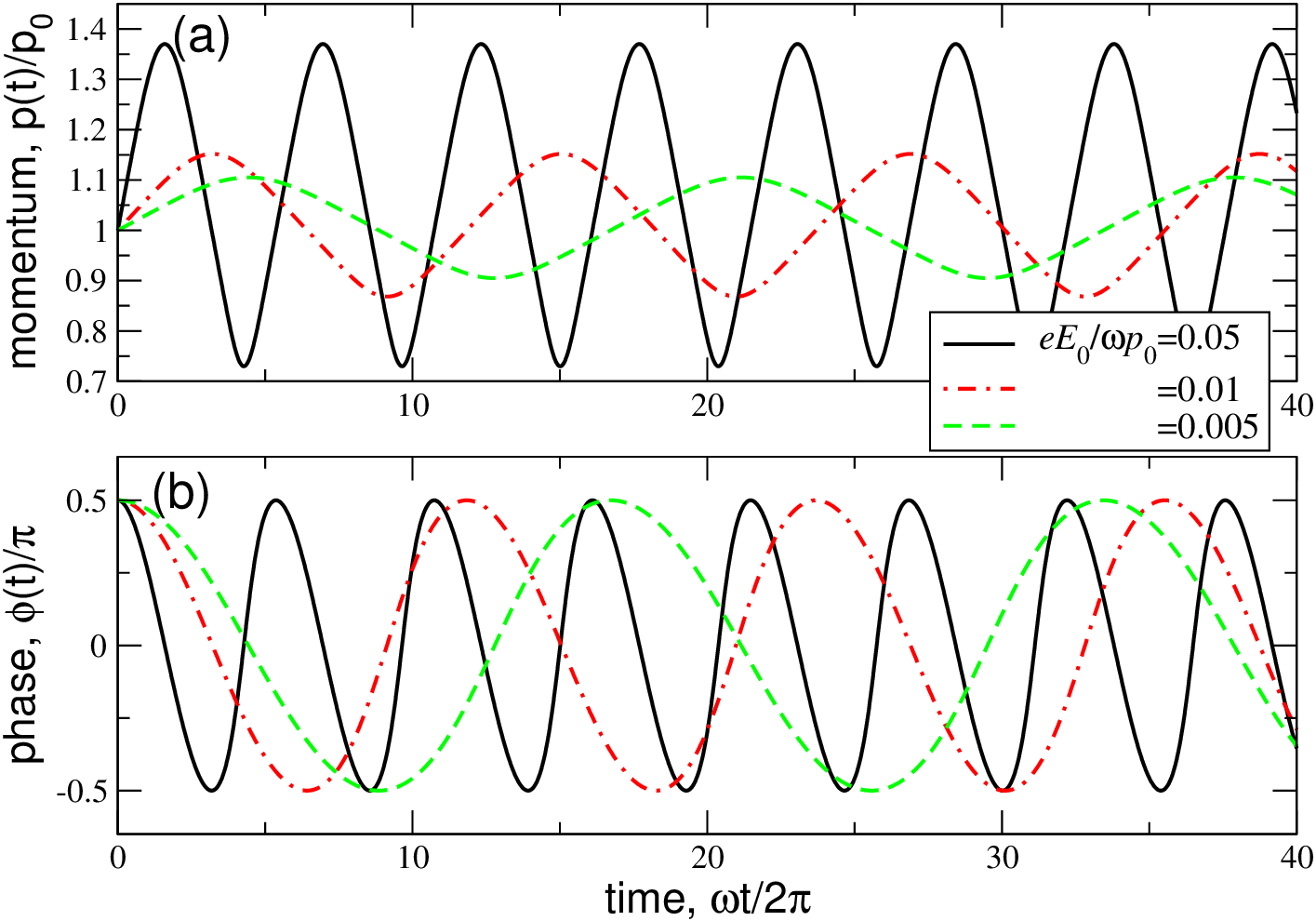}\hfill
\includegraphics[width=8.5cm]{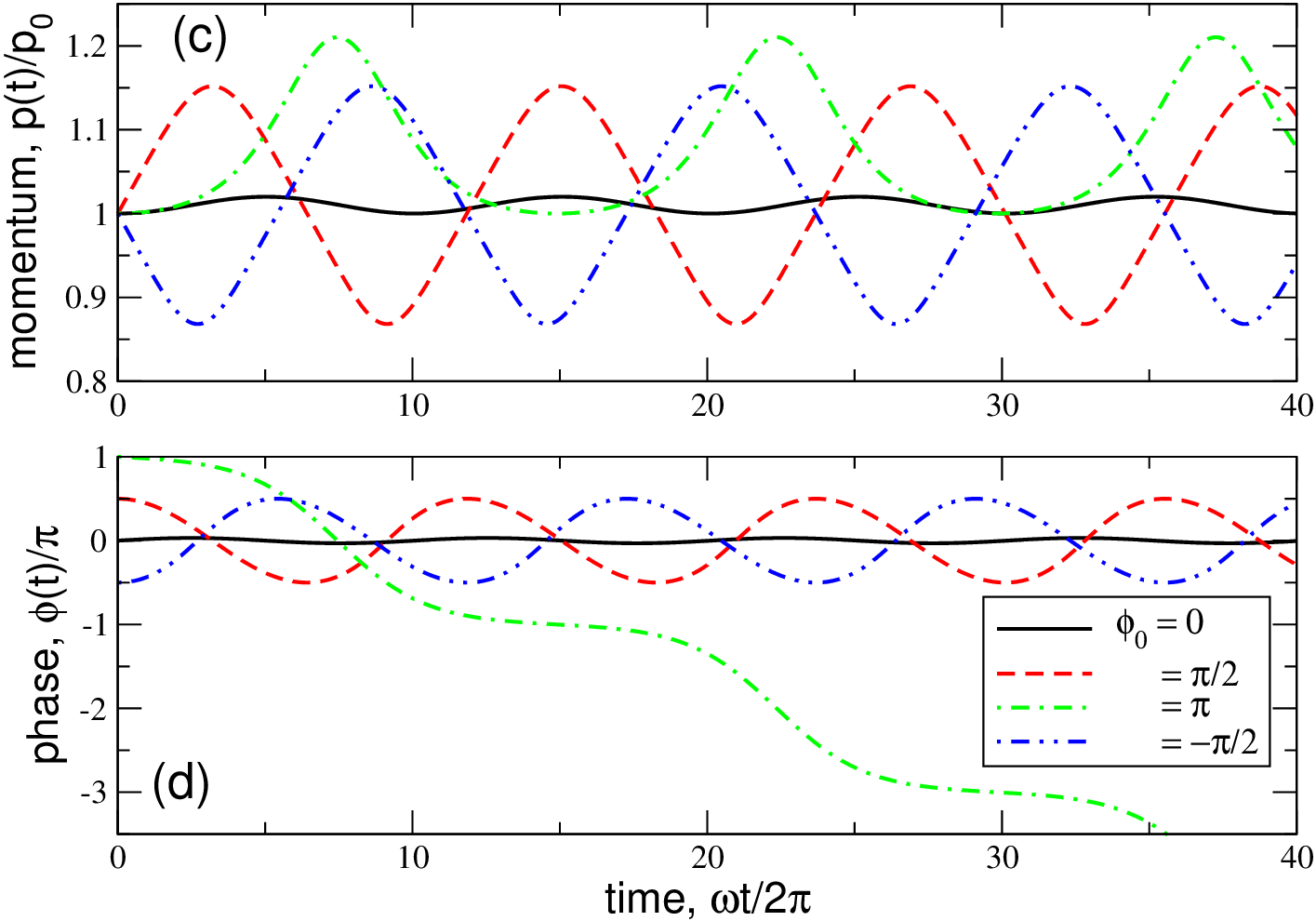}
\caption{\label{modphi} (Color online) Time dependencies of the momentum $p(t)$ and the phase $\phi(t)$ at different values of the electric field {\bf (a,b)} and under different initial conditions $\phi|_{t=0}=\phi_0$ {\bf (c,d)}. 
}
\end{figure}

To study this phenomenon further we introduce, instead of $p_x(t)$ and $p_y(t)$, the new variables $p(t)$ and $\phi(t)$ according to the  formulas
\be
\left(
\begin{array}{l}
p_x(t) \\ p_y(t)
\end{array}
\right)=
p(t)\left(
\begin{array}{r}
-\sin(\omega t+\phi(t)) \\
\cos(\omega t+\phi(t))
\end{array}
\right)
\ee
and investigate the time dependencies of the momentum $p(t)$ and the phase $\phi(t)$. Equations (\ref{Newton}) are then rewritten as
\be
\dot p(t)=eE_0\sin\phi(t), 
\label{Newton1a}
\ee
\be
p(t)\dot\phi(t)=-\omega p(t)+eVB/c+eE_0\cos\phi(t),
\label{Newton1b}
\ee
and the initial conditions are $p|_{t=0}=p_0$, $\phi|_{t=0}=\phi_0$. Figures \ref{modphi}a,b show the time dependencies of $p(t)$ and $\phi(t)$ at different values of the electric field parameter ${\cal F}=eE_0/\omega p_0$. Both functions are modulated, and the modulation frequency, as well as the amplitude of the momentum oscillations, decrease with ${\cal F}$. In contrast, the amplitude of the phase oscillations remains independent on the electric field $E_0$, Fig. \ref{modphi}b.  

The behavior of the $p(t)$ and $\phi(t)$ oscillations also depends on the initial phase $\phi_0$, Fig. \ref{modphi}c,d. One sees that not only the oscillation period and amplitude may depend on  $\phi_0$, but even the overall shape of oscillations, see the case $\phi_0=\pi$ (green dash-dotted curve on Fig. \ref{modphi}c,d). The linear contribution to the phase, Fig. \ref{modphi}d, means the frequency shift, since $\phi(t)=-\alpha t+\tilde\phi(t)$ leads to $\omega t+\phi(t)=(\omega-\alpha)t +\tilde\phi(t)$.  

\begin{figure}
\includegraphics[width=8.5cm]{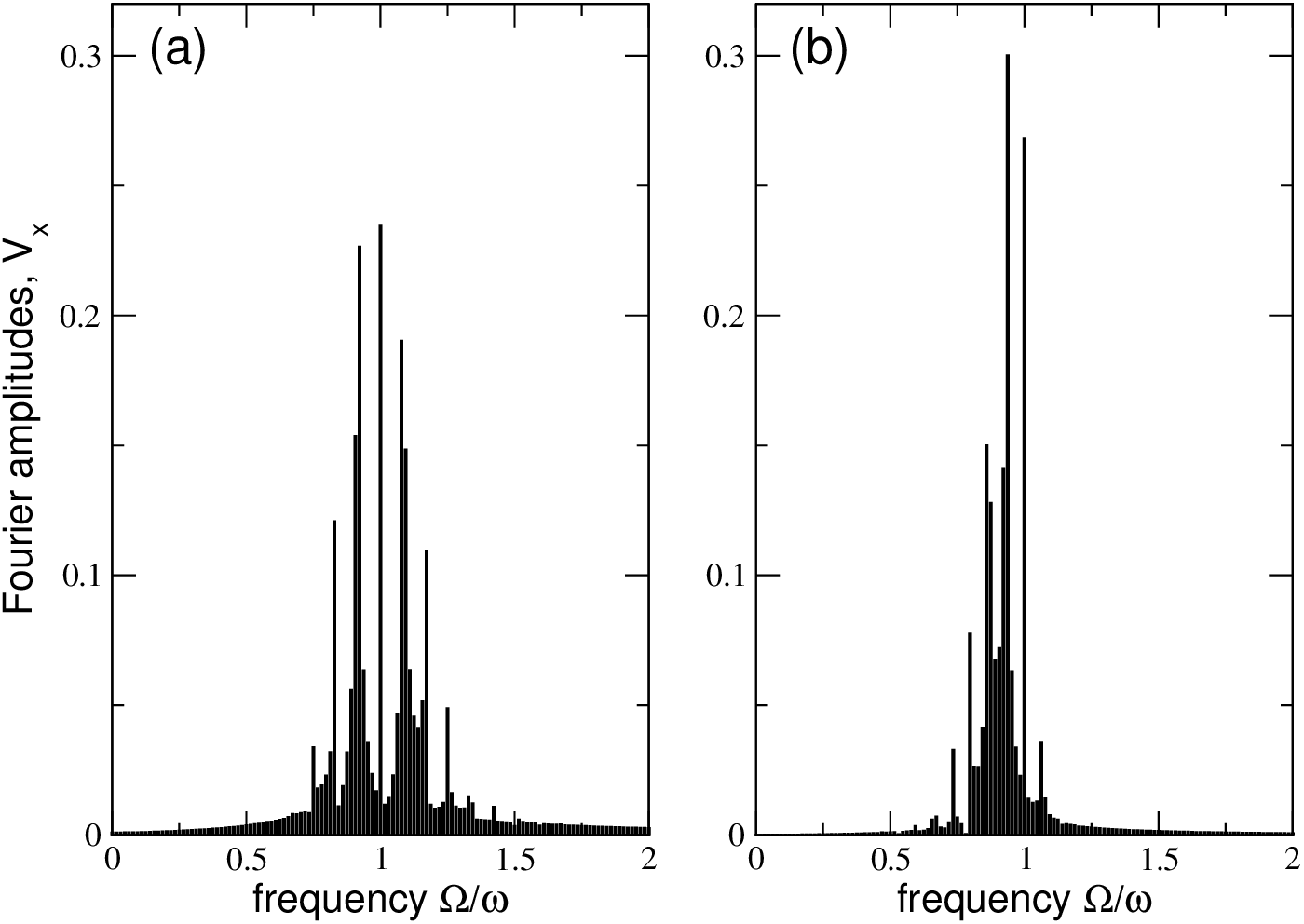}
\caption{\label{FFT} Frequency spectra of the velocity $v_x$ for ${\cal F}=eE_0/\omega p_0=0.01$ and $\phi_0=\pi/2$ {\bf (a)} and $\phi_0=\pi$ {\bf (b)}.
}
\end{figure}

So far we discussed the behavior of the momentum of the massless quasi-particle. The experimentally measured value, however, is the velocity 
\be
{\bf v}(t)=V\frac{{\bf p}(t)}{p(t)}=V\left(
\begin{array}{r}
-\sin(\omega t+\phi(t)) \\
\cos(\omega t+\phi(t))
\end{array}
\right).\label{v}
\ee
Its absolute value $|{\bf v}(t)|=V$ remains constant independent on how big or small the particle momentum is. This means that, even in a weak external electric field $E_0$ which produces small momentum oscillations, the induced ac current ${\bf j}(t)=-en_s{\bf v}(t)$ (in a system of many massless quasi-particles) can be rather strong.  For example for a typical electron density of $n_s=10^{12}$ cm$^{-2}$ in graphene the estimate gives a very big value of the current $j\approx 16$ A/cm. Another important feature of the time dependent velocity (\ref{v}) is that its phase remains to be modulated even in the limit of week electric fields ${\cal F}\to 0$, Fig. \ref{modphi}b. The Fourier spectrum of the velocity, Figure \ref{FFT}, shows therefore a broad spectrum around the central frequency $\Omega=\omega$ at ${\cal F}=eE_0/\omega p_0\ll 1$. Since the velocity of particles determines the current and the amplitudes of the scattered electromagnetic waves, Fig. \ref{FFT} shows, in  fact, the Fourier spectrum of the scattered waves $S_{scat}(\Omega)$. The function $S_{scat}(\Omega)$ contains a lot of Fourier harmonics although the spectrum of the incident wave is proportional to $\delta(\Omega-\omega)$. We emphasize that we discuss the CR resonance of the massless quasi-particle {\em in the absence of any scattering}.

In the limit ${\cal F}\ll 1$ one can get some more results analytically. Substituting $p(t)=p_0[1+q(t)]$ with $|q(t)|\ll 1$, we rewrite Eqs. (\ref{Newton1a})--(\ref{Newton1b}) as 
\be
\dot{q}(t)=\omega{\cal F}\sin\phi,\ \ \ \ 
\dot{\phi}(t)=-\omega q+\omega{\cal F}\cos\phi.\label{fulleq}
\ee
It will be seen from the result (\ref{res}) that $q\propto \sqrt{{\cal F}}$, therefore at ${\cal F}\ll 1$ one can neglect the last term in the second equation (\ref{fulleq}) as compared to the first one. Then the problem  (\ref{fulleq}) is reduced to the equation of the non-linear pendulum 
\be
\ddot{\phi}(t)=-\omega^2{\cal F}\sin\phi.\label{þhi-eq}
\ee
If the initial phase $\phi_0$ is small, $\phi_0\lesssim 1$, the solution assumes the form 
\be
\phi(t)=\phi_0\cos(\omega\sqrt{{\cal F}}t), \ \ 
q(t)=\phi_0\sqrt{{\cal F}}\sin(\omega\sqrt{{\cal F}}t).
\label{res}
\ee
The amplitude and the phase of the momentum oscillations are thus modulated with the frequency $\omega\sqrt{\cal F}$ proportional to the square root of the electric field. 

Substituting the phase $\phi(t)$ from (\ref{res}) to (\ref{v}) one can calculate the Fourier spectrum of the velocity $\tilde {\bf v}(\Omega)=\int {\bf v}(t)e^{-i\Omega t}dt/2\pi$. For example, for the $v_y$ component we get
\be
\tilde v_y(\Omega)=
\frac V{2}
\sum_{k=-\infty}^\infty  
i^k J_k(\phi_0) \delta(\Omega-\omega-k\omega\sqrt{\cal F}) +
\frac V{2}
\sum_{k=-\infty}^\infty (-i)^k J_k(\phi_0)
\delta (\Omega+\omega -k\omega\sqrt{\cal F}).
\ee
The spectrum consists of an in infinite number of satellite harmonics at the frequencies $\Omega= \omega \pm k\omega\sqrt{\cal F}$ with $k=1,2,\dots$. The amplitudes of these harmonics $J_k(\phi_0)$ are determined only by the initial phase of the particle and do not depend on the electric field. 

\begin{figure}
\includegraphics[width=8.5cm]{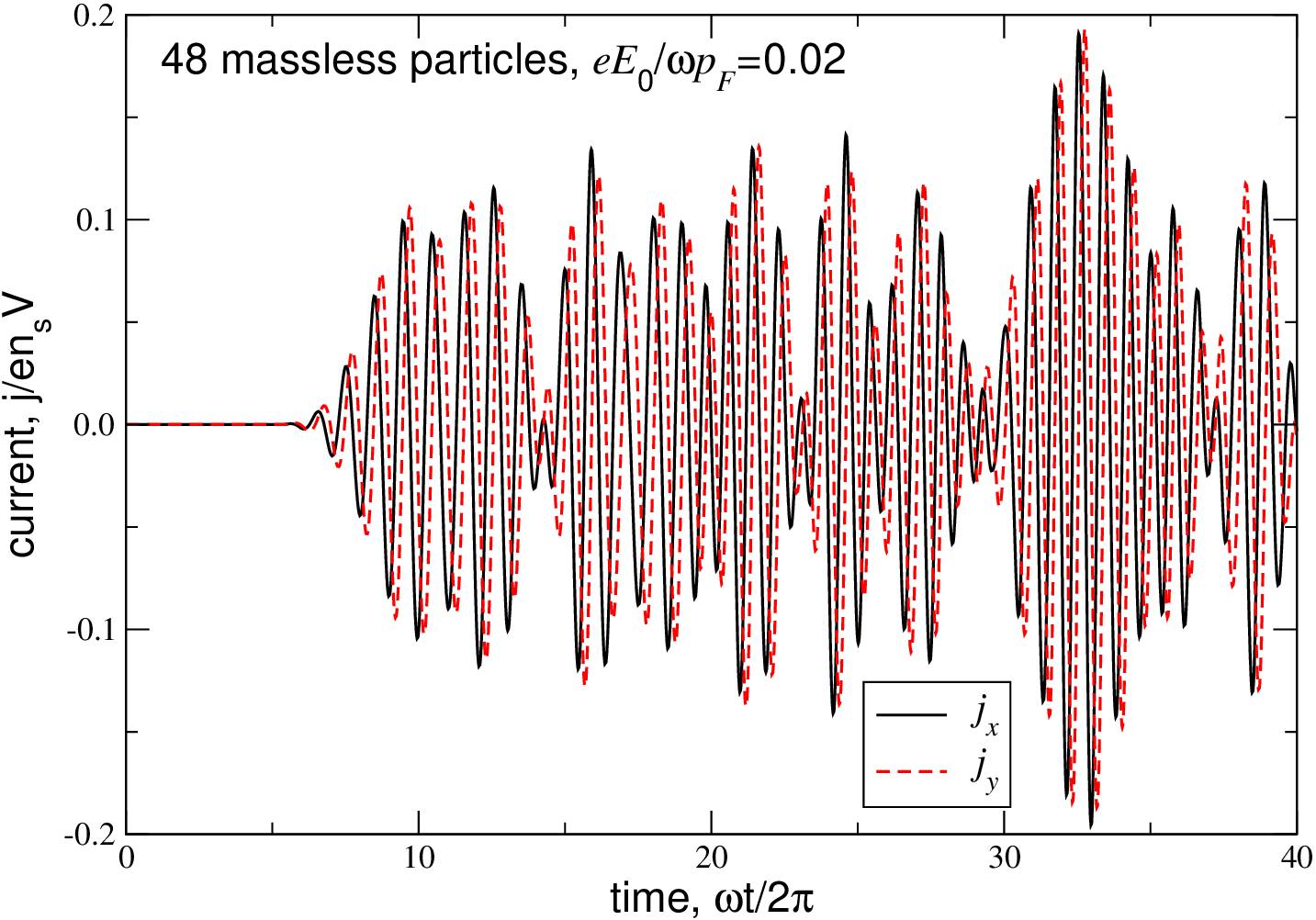}
\caption{\label{48p} (Color online) Time dependence of the current in a system of 48 massless particles. It is assumed that the particles are Fermi-distributed in the ${\bf p}$-space at $t=0$ and that the electric field is zero at $\omega t/2\pi<5$ and smoothly grows up to a constant value at $5<\omega t/2\pi<10$. 
}
\end{figure}

So far we discussed the response of only one quasi-particle and found that it is complicated and very different even for particles with the same $p_0$ but different initial phases $\phi_0$. In a real graphene system one deals with many particles with different initial phases and initial momenta $p_0\lesssim p_F$ lying inside the Fermi circle. The electromagnetic response of the whole many-particle system will be a superposition of partial responses of all the individual particles and will therefore be very complicated. The time dependence of the current (see example in Figure \ref{48p}) looks very chaotically, and the scattered wave spectrum $S_{scat}(\Omega)$ will have a broad peak around $\Omega=\omega$ with the quality factor only slightly exceeding unity. Such a large linewidth of the scattered wave resonance is not related to the scattering processes but is due to the linear graphene dispersion (\ref{disp}). Notice that the very broad CR lines have been observed in the experiments \cite{Jiang07,Deacon07}. As follows from our results they are not necessarily related to disorder: The broad CR line can be observed even in a perfectly pure graphene. It should also be noticed that chaotic motion of graphene quasi-particles has been discussed in other respects in Refs. \cite{Schliemann08,Rusin08} and the non-linear features of the CR in systems with the non-parabolic energy dispersion have been studied in \cite{Kaplan82}.

Above we assumed that the external ac electric field is {\em weak}, ${\cal F}\ll 1$. In such cases one usually uses the linear response theory which automatically implies that the system responds at the frequency coinciding with the frequency of the external radiation. As we have shown, this is not the case for the massless quasi-particles with the spectrum (\ref{disp}).  It should therefore be concluded that the current-vs-field dependence in graphene is non-analytical at low electric fields,  and that {\em the linear response theory does not adequately describe the electromagnetic response of graphene in finite magnetic fields}. 

Mathematically, the non-analyticity of the graphene response function originates from the singular Lorentz-force term in the equation of motion (\ref{Newton}). This singularity disappears at $B=0$, therefore in zero magnetic field the linear-response theory should be valid. This agrees with our previous results from Refs. \cite{Mikhailov07e,Mikhailov08a} where the non-linear frequency multiplication effect, predicted at ${\cal F}\gg 1$, was shown to disappear in the weak fields.  
 
To summarize, electromagnetic response of graphene in a magnetic field was shown to be essentially non-linear. The system irradiated by the monochromatic wave scatters the radiation in a broad frequency range. In agreement with experiments, the CR line was shown to be very broad in graphene, even if the intensity of the incident radiation is weak and all scattering processes are neglected. The linear response theory was shown to be not completely adequate for the  description of electromagnetic properties of graphene. Instead, the methods of the non-linear dynamics and the theory of chaos should be used for this purpose.

I thank Timur Tudorovskiy, Igor Goychuk, Vladimir Sablikov, Ulrich Eckern and Levan Chotorlishvili for useful discussions. The work was supported by the Swedish Research Council and INTAS.


\end{document}